\documentclass[conference]{llncs}

\usepackage{url}
\usepackage{cite}
\usepackage{graphicx}
\usepackage{amsmath}
\usepackage{subfigure}
\usepackage[ruled,figure]{algorithm2e}

\urldef{\mailsa}\path|{vss@iisc.ac.in}|

\begin{document}

\title{Strategies for Efficient Executions of Irregular Message-Driven Parallel Applications on GPU Systems}

\author{Vasudevan Rengasamy \and Sathish Vadhiyar}
\institute{Department of Computational and Data Sciences, Indian Institute of Science, Bangalore, India\\
\mailsa\\
}

\maketitle

\begin{abstract}
Message-driven executions with over-decomposition of tasks constitute an important model for parallel programming and have been demonstrated for irregular applications. Supporting efficient execution of such message-driven irregular applications on GPU systems presents a number of challenges related to irregular data accesses and computations. In this work, we have developed strategies including coalescing irregular data accesses and combining with data reuse, and adaptive methods for hybrid executions to minimize idling. We have integrated these runtime strategies into our {\em G-Charm} framework for efficient execution of message-driven parallel applications on hybrid GPU systems. We demonstrate our strategies for irregular applications with an N-Body simulations and a molecular dynamics application and show that our dynamic strategies result in 8-38\% reduction in execution times for these irregular applications over the corresponding static strategies that are amenable for regular applications.
\end{abstract}

\section{Introduction}

Message-driven executions with over-decomposition of tasks constitute an important model for parallel programming and provides multiple benefits including providing communication-computation overlap and minimizing idling on resources.  \textsc{Charm++} \cite{kale-charm-pp1996} from UIUC, USA is one such message-driven programming environment and runtime for parallel applications.  \textsc{Charm++} has been used to provide high performance for different scientific applications including NAMD \cite{mei-enablingandscaling-sc2011}, a  molecular dynamics application, ChaNGa \cite{jetley-massivelyparallel-ipdps2008}, a cosmological simulator and ParFUM \cite{lawlor-parfum-ec2006}, a framework for unstructured mesh applications.

In our earlier work \cite{vasudevan-gcharm-ics2013}, we had developed an adaptive runtime framework called {\em G-Charm} for efficient executions of \textsc{Charm++} message-driven parallel applications on hybrid GPU systems. Our framework dynamically combined multiple kernels corresponding to large number of small \textsc{Charm++} objects ({\em chares}) into a single GPU kernel to reduce the number of kernel invocations, managed data movement between CPU and GPU, provided reuse of data on GPUs and avoided redundant CPU-GPU data transfers by analyzing dependencies, and enabled dynamic scheduling of tasks for asynchronous executions on both CPUs and GPUs using performance estimates.

The techniques in our framework were amenable mostly for regular applications like matrix computations.
Supporting efficient executions of message-driven irregular applications on GPU systems presents a number of challenges to our G-Charm runtime system. The generation of the tasks (or chares) by the application need not be periodic. Our runtime framework needs to wait for sufficient number of tasks for combining the chares into a GPU kernel.
Accesses of the data by the tasks can be irregular. Multiple tasks may require overlapping sets of data located in different regions of memory. This presents a problem for our data reuse policies since avoiding transfers of data already located in GPU memory for a kernel can result in non-coalesced memory accesses since the data already present in GPU memory may be widely separated from the data that is to be transferred for the current kernel invocation. Finally, the strategies for asynchronous executions will have to adapt to the varying workloads of the tasks that are mapped to CPUs and GPUs.

In this work, we have developed strategies in G-Charm for efficient executions of irregular message-driven parallel applications on GPU systems. We have developed models for deciding the number of tasks to aggregate to a kernel based on the rate of task generation, and the GPU occupancy of the tasks. We have also developed a data reorganization strategy for improving coalesced memory access for irregular data accesses on GPU, and combines data coalescing with data reuse. Finally, our runtime framework dynamically adapts hybrid executions on CPUs and GPU to the amount of computations and communications in the tasks. We demonstrate our strategies for irregular applications for ChaNGa N-Body simulations application \cite{jetley-massivelyparallel-ipdps2008} and a molecular dynamics (MD) application, and show that our dynamic strategies result in 8-38\% reduction in execution times for these irregular applications over the corresponding static strategies that are amenable for regular applications. 

\section{Background}
\label{bg}

\subsection{\textsc{Charm++}}

\textsc{Charm++} is a message-driven object oriented parallel programming framework based on C++ \cite{kale-charm-pp1996}. A parallel application written using \textsc{Charm++} divides the data among an array of objects called {\em chares}. The chares are mapped to physical processors, and can be migrated among processors by the \textsc{Charm++} runtime system to provide load balance. Typically, the number of chares are much larger than the number of physical processors, resulting in over-decomposition. The chares are associated with specialized methods called {\em entry} methods. Entry methods of a chare object can be invoked from chares present in same or other processors. Remote entry methods invoked by a chare are queued as messages in a message queue at the destination processor. The runtime system dequeues a message and invokes the corresponding chare's entry method upon arrival of all inputs of the entry method from the other chares. Thus, while input data for a chare is communicated from a remote processor, a processor can perform computation on some other chare for which inputs have already arrived.

\subsection{\textsc{G-Charm}}

In our earlier effort \cite{vasudevan-gcharm-ics2013}, we had developed G-Charm runtime framework that performs various optimizations including minimizing data transfers, data management, dynamic scheduling and work agglomeration. The runtime automatically decides the allocation of some of the CPU chares for execution on the GPU, and converts the chares to {\em GPU chares}. The runtime is also responsible for transferring data to GPU, invoking kernels, periodically monitoring the status of kernels, copying data to CPU upon kernel completion, and invoking a callback function on CPU to notify chares about the work completion.
The application begins execution with the creation of chare objects. Each chare operates on a subset of data and executes its entry methods to update its own data on the arrival of input data from other chares and also to invoke entry methods of other chares. When a chare needs to invoke a kernel on the GPU, it creates a {\em workRequest} object and invokes a scheduler function in G-Charm runtime that performs dynamic scheduling of the workRequest to either CPU or GPU.
The G-Charm runtime checks the data region in the application domain represented by the workRequest data, and tries to avoid redundant data transfers to GPU by transferring only the data not already present in GPU. The G-Charm runtime then adds the workRequest to a node of a linked list called {\em workGroupList}, in which each node represents a set of workRequest objects that can be combined. G-Charm periodically combines workRequest objects from this list and creates objects of type {\em workRequestCombined}. The G-Charm runtime then schedules these objects for GPU execution. Thus the G-Charm runtime performs work agglomeration dynamically by combining kernels of multiple work requests for GPU execution.

\section{Optimization Strategies for Irregular Applications}
\label{opt}

\subsection{Combining Kernels}
\label{combining}

Combining multiple kernels of different chares into a single large kernel results in smaller number of kernel invocations, smaller CPU-GPU data transfer costs and larger GPU occupancy. Our G-Charm framework dynamically selects the workRequest objects of different chares for combining into a single kernel.
For irregular applications, the arrival rate of the workRequests or tasks to the workGroupList for combining can vary throughout application execution. After a fixed interval when the {\em combine} routine is called, if the workGroupList has only a small number of workRequests to combine, the resulting GPU kernel can be spawned with only a small number of threads and thread blocks, thus resulting in poor GPU occupancy. However, waiting for a fixed number of workRequests to arrive in the workGroupList before spawning the GPU kernel for good GPU occupancy can result in large idling of the GPU if the workRequests do not arrive within a guaranteed period of time.

Thus, a strategy for irregular applications has to consider both the arrival rate and the GPU occupancy to decide on combining the workRequests into a kernel. In our work, our G-Charm runtime system uses the CUDA occupancy calculator to determine the percentage occupancy and the maximum number of thread blocks that can be used per Streaming Multiprocessor (SM) to achieve the occupancy for a kernel.
Since in our implementation, a workRequest is executed using a thread block, the maximum number of thread blocks obtained from the CUDA occupancy calculator also corresponds to the maximum number of workRequests, $maxSize$, that can be combined for maximum occupancy.

Our runtime also notes the times of workRequest generation or arrival, and maintains a running maximum of the intervals, $maxInterval$, between the arrivals using these timestamps. Our framework periodically checks the workGroupList. If the number of workRequests in a workGroupList is at least $maxSize$, then it combines $maxSize$ number of workRequests into a combined kernel for GPU execution. If the number is less than the $maxSize$, G-Charm finds the interval between the current time and the time when the last workRequest arrived. If this interval is greater than $2\times maxInterval$, it combines the available workRequests for immediate execution. Thus, our framework attempts to achieve a balance between providing maximum GPU occupancy and minimum GPU idling.

\subsection{Data Reuse and Coalescing}
\label{re-co}

Data transfers on the PCI/e bus between CPU and GPU for kernel executions can occupy significant times in overall execution. Hence it is important to minimize these times by avoiding the transfer of some of the data for a given kernel execution if the data is already located in the GPU memory due to previous kernel executions. G-Charm keeps track of the data segments in the GPU device used for kernel executions. Each chare is associated with a region of data in the application domain.
The G-Charm runtime keeps track of the mapping of chare buffers to slots in the device memory using a {\em chare table}.
A workRequest object contains the indices of the chare buffers representing subregions in the application domain. When a workRequest for a chare is created, the G-Charm runtime uses the buffer indices of the workRequest to lookup the chare table and find if the buffers are already located in the GPU memory due to the prior execution of kernels of other chares on the GPU (e.g., data generated from previous iterations).

While minimizing the transfers is important, it is also essential to provide data locality of the data that is being reused. Data locality in the GPU memory results in coalesced access in which the data needed by the consecutive threads of a half warp (16 threads) are located in contiguous locations of the GPU device memory. Providing coalesced access in GPUs has consistently been shown as an important optimization providing large scale performance benefits.

\begin {figure}
\centering
\subfigure[Input Data and GPU State]{
\includegraphics[scale=0.28]{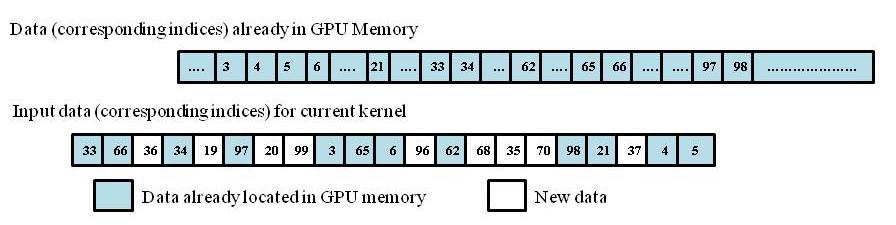}
\label{input_data}
}
\subfigure[Redundant Data Transfers and Coalesced Access] {
\includegraphics[scale=0.28]{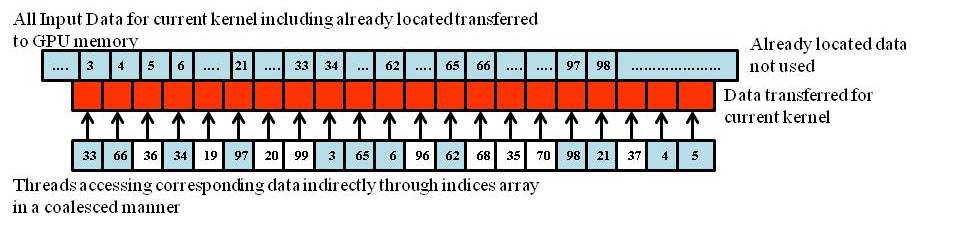}
\label{redundant}
}
\subfigure[Data Reuse, Uncoalesced Access] {
\includegraphics[scale=0.28]{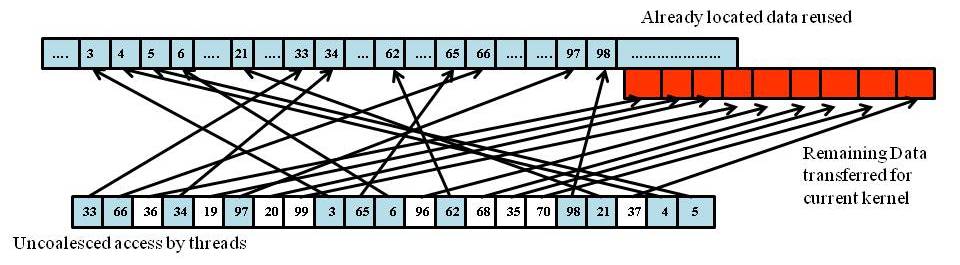}
\label{reuse_uncoalesced}
}
\subfigure[Data Reuse, Sorting Indices, Coalesced Access] {
\includegraphics[scale=0.28]{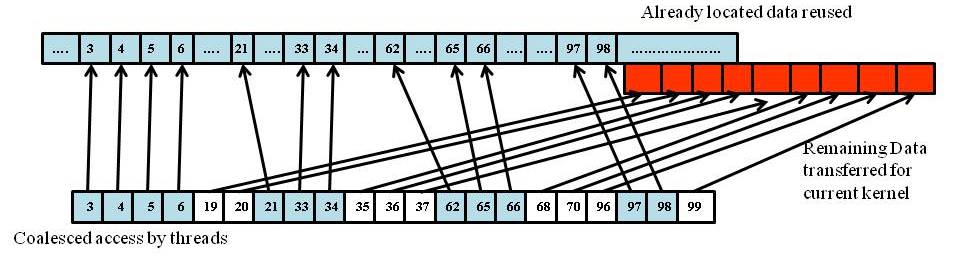}
\label{reuse_coalesced}
}
\caption{Data Reuse and Coalescing Strategies}
\label{reuse-coalescing}
\end{figure}

Irregular parallel applications exhibit poor data locality. In these applications, reuse of data already located in the GPU device memory will impact the locality of the data needed by consecutive threads: if the data required by thread $t_i$ is already located in GPU device memory at location $index_i$, then the data needed by thread $t_{i+1}$ may not already be located in location  $index_i+1$, but may either be already located in some other location or may not be located at all. 
Thus reuse of data can significantly upset the coalesced memory access in irregular applications. In extreme cases, the gain obtained due to minimizing data transfers by reusing data can be offset by the loss in performance due to non-coalesced memory access. In these cases, it may be better to not reuse data, but perform redundant transfers of all the data needed by the current kernel such that the new data is organized for coalesced access. This is illustrated in Figure \ref{redundant}. The figure shows the redundant data transfers corresponding to the input data for the current kernel and the current GPU state shown in Figure \ref{input_data}. Figure \ref{reuse_uncoalesced} shows the non-coalesced access that can happen in irregular applications due to data reuse.

Thus, for irregular applications, data reuse optimization has to be followed by a data reorganization or task reassignment step for coalesced access. Data reorganization for coalesced access in irregular applications is challenging. 
We follow a simple strategy in which we reassign the tasks to the threads. This is done by sorting the indices of the data accessed by the threads and making the threads access the data in the sorted order of the indices. This results in local sets of contiguous data accesses as illustrated in Figure \ref{reuse_coalesced}. As our results in Section \ref{exp} show, this simple reassignment achieves significant improvements in performance of kernel execution.

For sorting the data indices for coalesced access, one method is to perform the sorting of the indices array after combing the workRequests. However, this will introduce high sorting overheads of $O(N^2)$ or $O(Nlogn)$ for $N$ data items. Instead, given a sorted sub-array corresponding to the earlier workRequests, G-Charm inserts an index for a data item corresponding to the current workRequest in the correct position during the invocation of {\em gcharm-insertRequest()} function such that the resulting expanded array continues to maintain the sorted order. The correct position for insertion of the data item into the array is performed using binary search. The complexity of this will be $O(log1+log2+log3+...+logN)=O(log(N!))$.

\subsection{Dynamic Scheduling}
\label{dyn}

G-Charm dynamically decides the allocation of a chare to either CPU or GPU for tasks for which kernel functions exist for both CPU and GPU.
G-Charm performs dynamic scheduling of tasks of a kernel or functions for asynchronous executions on CPU and GPU cores by executing the initial tasks on both CPU and GPU, obtain the performance ratio of executions, and use this ratio as estimates for subsequent tasks.
We model the workload of a workRequest based on the amount of input data accessed by the workRequest. This information of the data buffer indices accessed by a workRequest is maintained for the data reuse optimization described in the earlier section. After every execution of a combinedWorkRequest on a CPU or GPU, our framework obtains the times taken for execution per input data item in the workRequest on both the devices. These times are dynamically updated as running averages of the times obtained till the current point in the execution. The performance ratio between the CPU and GPU times per data item is calculated.

Given a queue of workRequests, first the total number of data items across all the workRequests in the queue is found. The total number is divided using the performance ratio between CPU and GPU to find the number of data items that have to be allocated to the CPU and GPU. The workRequests are then scanned from the beginning of the queue, and a running cumulative sum of the number of data items in the workRequests scanned is maintained. If this cumulative sum crosses the number of data items to be allocated to CPU, the set of workRequests scanned so far are allocated to CPU and the remaining workRequests are allocated to GPU for execution. Thus, by considering the individual workloads of the workRequests and updating the performance ratios by maintaining running averages of the times per data item, our framework adapts to changing workloads for dynamic scheduling in irregular applications.

\section{Experiments and Results}
\label{exp}

We demonstrate our techniques for ChaNGa cosmology simulations application \cite{jetley-massivelyparallel-ipdps2008} and a molecular dynamics simulation application. We have performed single node experiments on two systems: one with a 6-core Intel Xeon E5-2620 processor connected to a single Kepler K20C GPU and the other with dual 8-core Intel Xeon E5-2670 processors connected to two Kepler K20m GPUs. 

\subsection{ChANGa N-Body Simulations Application}

ChaNGa is an iterative N-Body simulations application and uses parallel Barnes-Hut algorithm for calculation of interactions between the bodies. In ChaNGa, particles are divided among {\em TreePiece} chares with each chare representing a part of the Barnes-Hut tree. Each iteration involves domain decomposition of particle space, distributed Barnes-Hut tree construction, local and remote tree walks to create interaction lists, gravitational force computation on particles due to interaction with tree nodes and other particles, force computations with periodic boundary conditions using Ewald summation, acceleration and updates of coordinates of particles. Particles are grouped into {\em buckets} and all particles in a bucket interact with same nodes and particles. 

Force computations and Ewald summation are done primarily on GPU. We use the GPU parallelization scheme of force computations by Jetley et al. \cite{jetley-scalinghierarchical-sc2010}. We use a 2D CUDA block of size $16\times8$ to compute force on each bucket. The threads in column $0$ load the bucket particles into shared memory. Threads in row $0$ load eight interactions into shared memory using which the threads in row $i$ compute force on particle $i$. Then the next set of interactions is loaded into shared memory and the process repeats until all interactions are completed.

We have used two datasets in our experiments.
\begin{itemize}
  \item cube300 - A low resolution cosmological simulation with $48^3$ particles in a cubic box of 300 Mpc per side. The application is executed with this dataset for 128 iterations.
  \item lambs: A larger dataset with $144^3$ particles in a cubic box of 71Mpc. The application is executed with this dataset for 10 iterations.
\end{itemize}
The particles in both datasets exhibit moderate clustering on small scale and become more uniformly distributed with increasing scale.

\subsection{Molecular Dynamics Simulation Application}

We consider a two-dimensional molecular dynamics application in which the 2D space is partitioned into patches. Each patch owns the particles present in the region. In each timestep, force on each particle due to other particles within a cutoff distance is calculated and the position of the particles are updated. Particles migrate to neighboring patches according to new positions and the application proceeds to next timestep. This is repeated for a fixed number of timesteps.

In the \textsc{Charm++} implementation, a {\em compute object} calculates force between a pair of patches. The entry method {\em interact} takes two vectors of particles belonging to two patches and updates force components of each particle. The widely-used NAMD \cite{mei-enablingandscaling-sc2011, phillips-adapting-sc2008} molecular dynamics framework based on \textsc{Charm++} also adopts a similar parallelization scheme based on compute objects. The {\em interact} method has been implemented as a CUDA kernel for the G-Charm implementation. 

\subsection{Combining Kernels}

Our framework calculated the GPU occupancy as 50\% and 31\% for the force computation and Ewald summation kernels, respectively. The maximum number of active blocks is 16 per SM (streaming multiprocessor) for the NVIDIA Kepler architecture. So the total number of blocks that can be active is 104 (8 blocks $\times$ 13 SMs for Kepler) for the force computation kernel and 65 ($4.8 \times 13$) for the Ewald summation kernel, respectively. Since each workRequest corresponds to one bucket and is executed with one CUDA block, the framework combines workRequests until the number of distinct buckets in the combinedWorkRequest is more than 104 in case of force computation kernel and 65 in case of Ewald summation kernel.

\begin {figure}
\centering
\subfigure[Small Dataset (cube300)]{
\includegraphics[scale=0.22, angle=270]{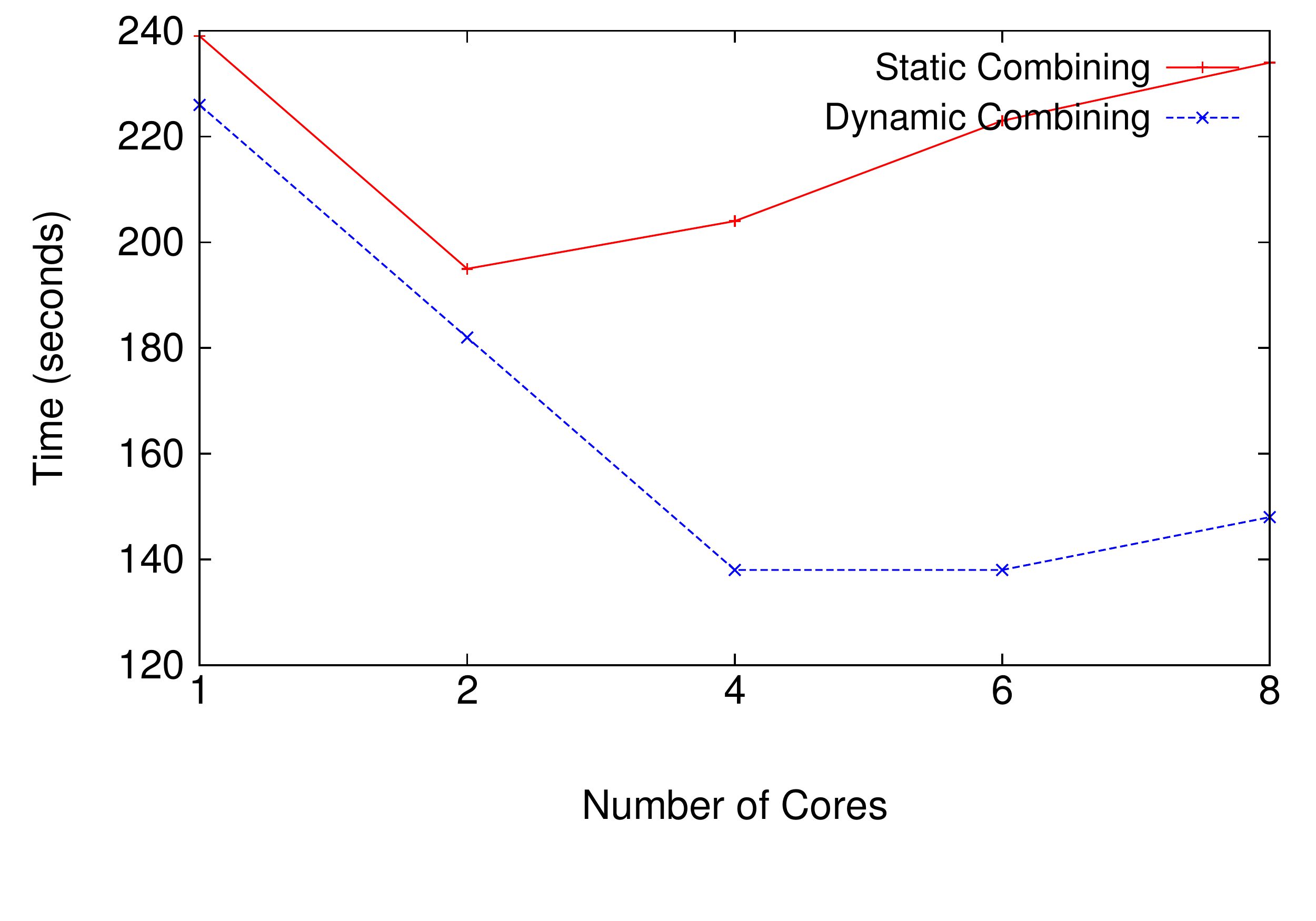}
\label{combine_cube300}
}
\subfigure[Large Dataset (lambs)] {
\includegraphics[scale=0.22, angle=270]{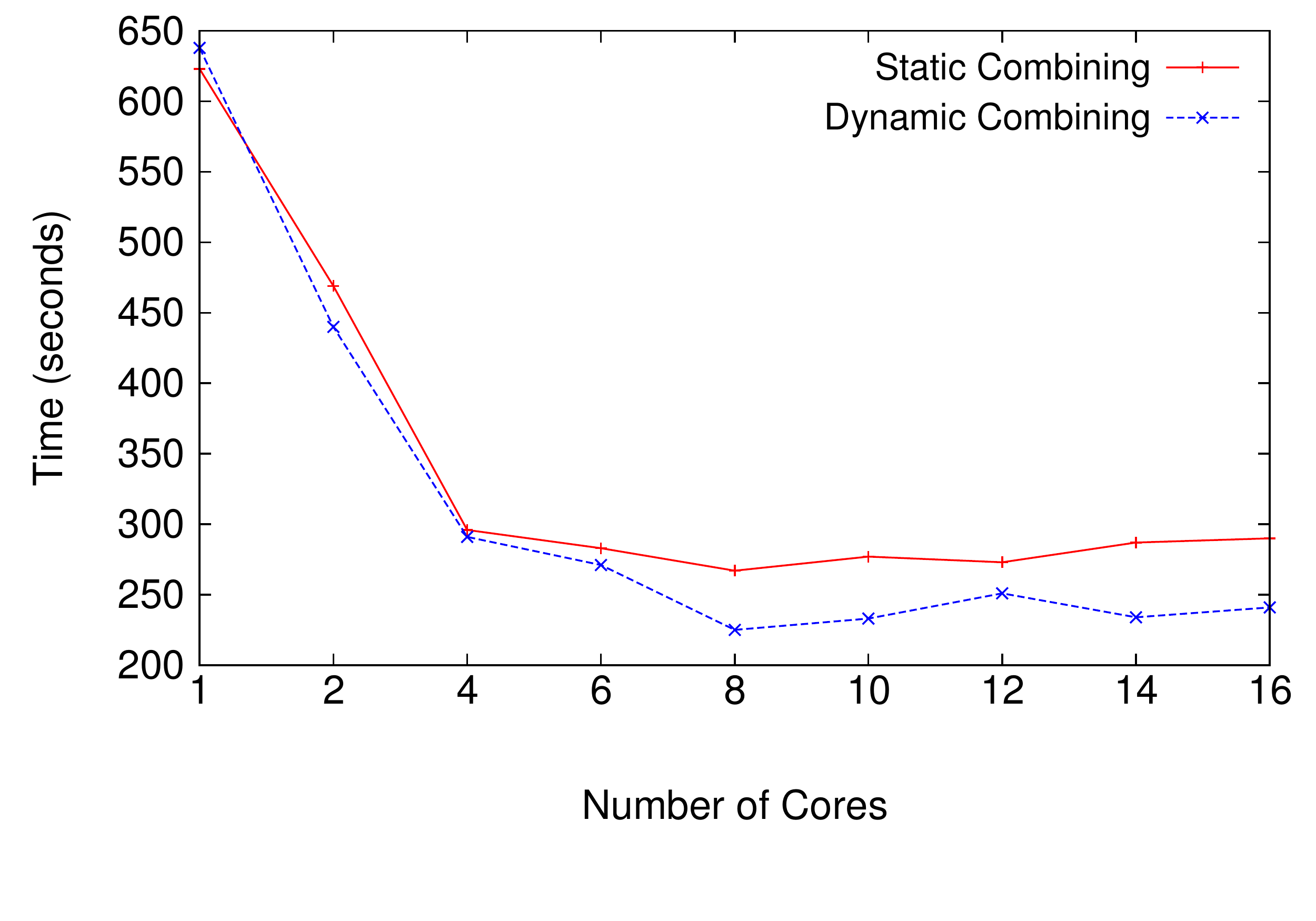}
\label{combine_lambs}
}
\caption{Dynamic vs Static Combining Strategies for Small and Large Datasets with ChaNGa}
\label{combining_results}
\end{figure}

Figure \ref{combining_results} shows the benefits of our strategy for combining small workRequests into a kernel for irregular applications. We compare our adaptive strategy that considers both the GPU occupancy and the arrival rate of the workRequests (Section \ref{combining}) with a static strategy that combines the available set of workRequests after processing every 100 workRequest objects in the CPU.
We find that the dynamic strategy gives about 8-38\% reduction in execution times over the static strategy for the small dataset, and about 19\% reduction for the large dataset.

\subsection{Data Reuse and Coalescing}

\begin {figure}
\centering
\includegraphics[scale=0.22, angle=270]{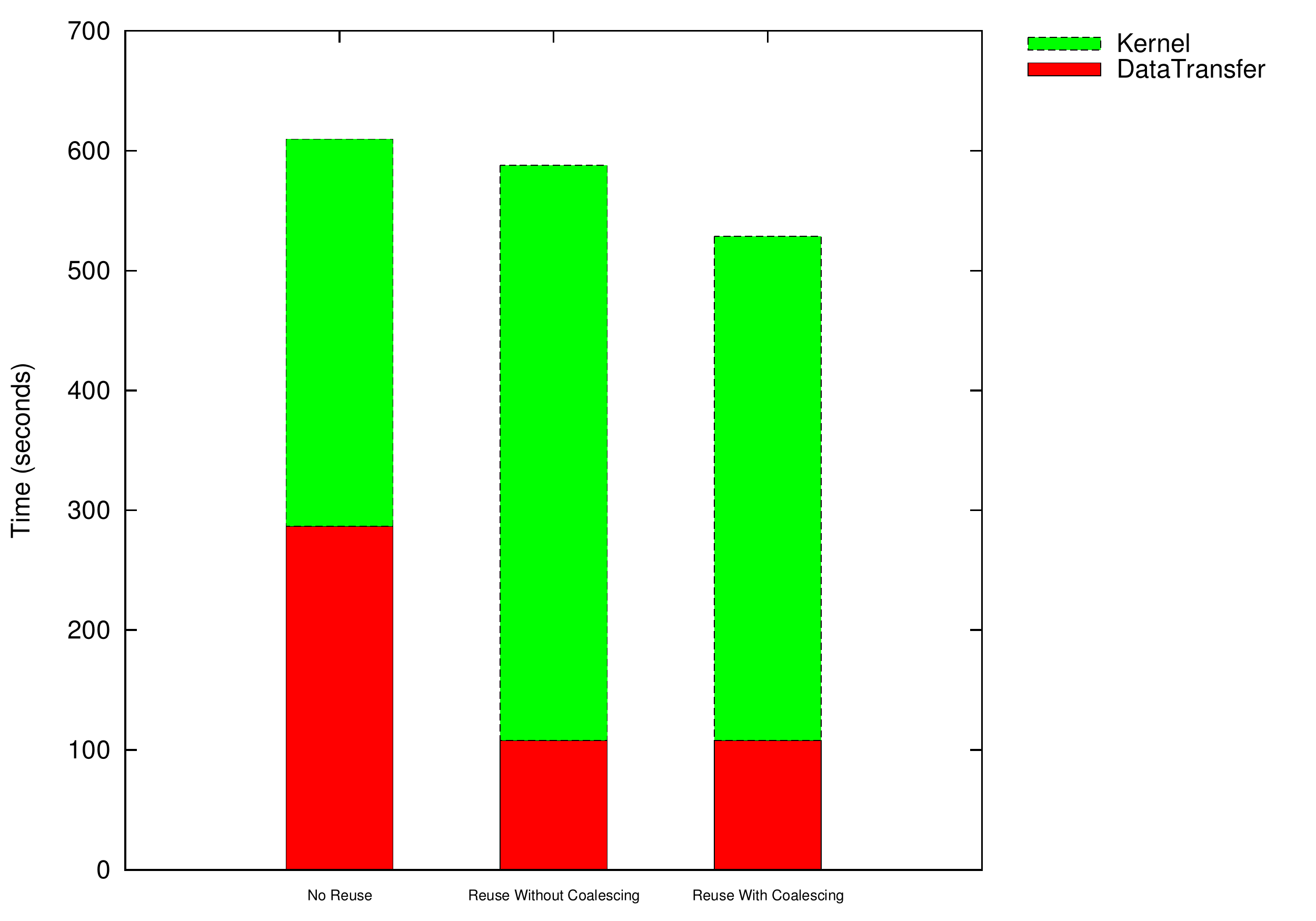}
\caption{GPU Kernel and Data Transfer Times for Large Dataset with ChaNGa on 8 Cores}
\label{reuse_coalescing_results}
\end{figure}

Figure \ref{reuse_coalescing_results} shows the GPU kernel times and the CPU-GPU data transfer times with redundant data transfers (no reuse), applying data reuse optimization, and applying both data reuse and improved coalesced access using sorting data indices (Section \ref{re-co}). The figure shows that applying only the reuse technique gives only 3.6\% reduction in execution time over the original code that employs redundant data transfers. 
Since data reuse results in non-contiguous memory accesses by members in a combinedWorkRequest object, an additional buffer containing the addresses of data items corresponding to individual workRequest objects has to be transferred to GPU before kernel invocation. This also doubles the number of accesses to global memory since for accessing each data item, the address has to be obtained from another buffer in global memory. While the time taken for data transfers reduces by 62\% due to the reduction in data transferred by employing data reuse, the GPU kernel time increases by 49\%. This is because of non-coalesced access of data already located in the GPU memory and the new data and the additional global memory accesses.

Combining reuse with coalesced access on GPU by sorting the array indices results in 12\% reduction in execution time over the original code with redundant data transfers and 8\% reduction in time over applying only the data reuse strategy. The coalesced access achieves about 10\% reduction in kernel execution time over applying only the reuse strategy that results in non-coalesced access. Note that the kernel computation is still higher than the original code, since the original code achieves complete coalescing, while our techniques of sorting array indices achieves only local regions of coalesced 
access as described in Section \ref{re-co}.

\subsection{Comparison with a Hand-Tuned Hybrid N-Body Simulations Code}

\begin{figure}
\centering
\includegraphics[scale=0.22, angle=270]{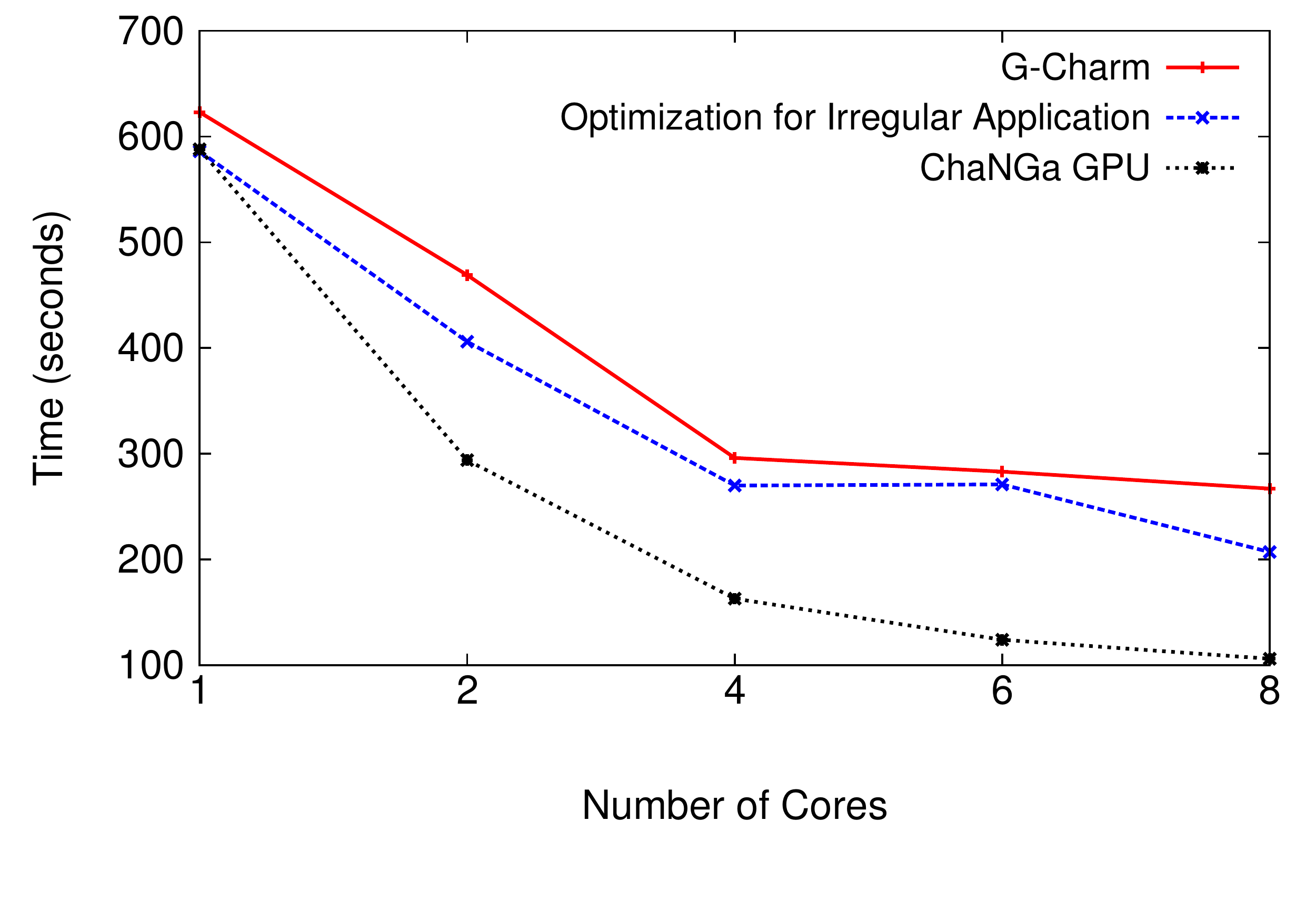}
\label{changa_comparison_lambs}
\caption{Comparison of Adaptive Strategies for Combining Kernels and Data Reuse with Static Strategies and a Hand-Tuned Code for ChANGa}
\label{changa_comparison}
\end{figure}

With our G-Charm adaptive strategies, we obtained about 31\% reduction in execution time for the cube300 dataset and about 62\% reduction in execution time for the lambs dataset, over average execution time of multi-core CPU implementation for up to 8 CPU cores. We also compared the total times for ChANGa N-Body simulations obtained using our adaptive strategies for combining, data reuse and coalescing described in this work with the total times obtained using the static strategies developed in our earlier work \cite{vasudevan-gcharm-ics2013} that are amenable for regular applications. We also compare with the results obtained with a hand-tuned version of ChANGa for the hybrid GPU architectures developed by Jetley et al. \cite{jetley-scalinghierarchical-sc2010}. This code was manually tuned by the developers for optimal data layout, hybrid executions and data transfers based on various parameter studies. In comparison, in our work, all these optimizations are performed automatically by the framework in generic ways without the knowledge of the application. Figure \ref{changa_comparison} shows the comparison results for the large dataset.

We see that our dynamic strategies perform better compared to static methods. Our methods show good scalability upto 8 cores and the scaling trend is similar to ChaNGa GPU implementation. The performance we obtain is less than ChaNGa GPU implementation mainly because of the overheads in our runtime system and lack of application specific optimizations such as the use of constant memory in ChaNGa GPU implementation to store the read only data for Ewald Kernel. But the strategies used in our runtime system are generic and can be applied to other irregular applications.

\subsection{Dynamic Scheduling}

Our profiling experiments with the ChANGa application showed that the CPU cores are sufficiently occupied with the tree traversal tasks. Hence there is no scope for applying our G-Charm's automatic dynamic scheduling techniques for asynchronous executions on CPU and GPU cores with this application.
We demonstrate our dynamic scheduling strategy for irregular applications using the molecular dynamics (MD) simulation application. The G-Charm framework automatically performs asynchronous computations of interaction calculations on both the CPU and the GPU cores. We compare our dynamic strategy that adapts to changing workloads of workRequests by considering the data accesses in the workRequests with a static strategy that partitions the workRequestQueue based only on the total number of workRequests in the queue.

\begin {figure}
\centering
\includegraphics[scale=0.22, angle=270]{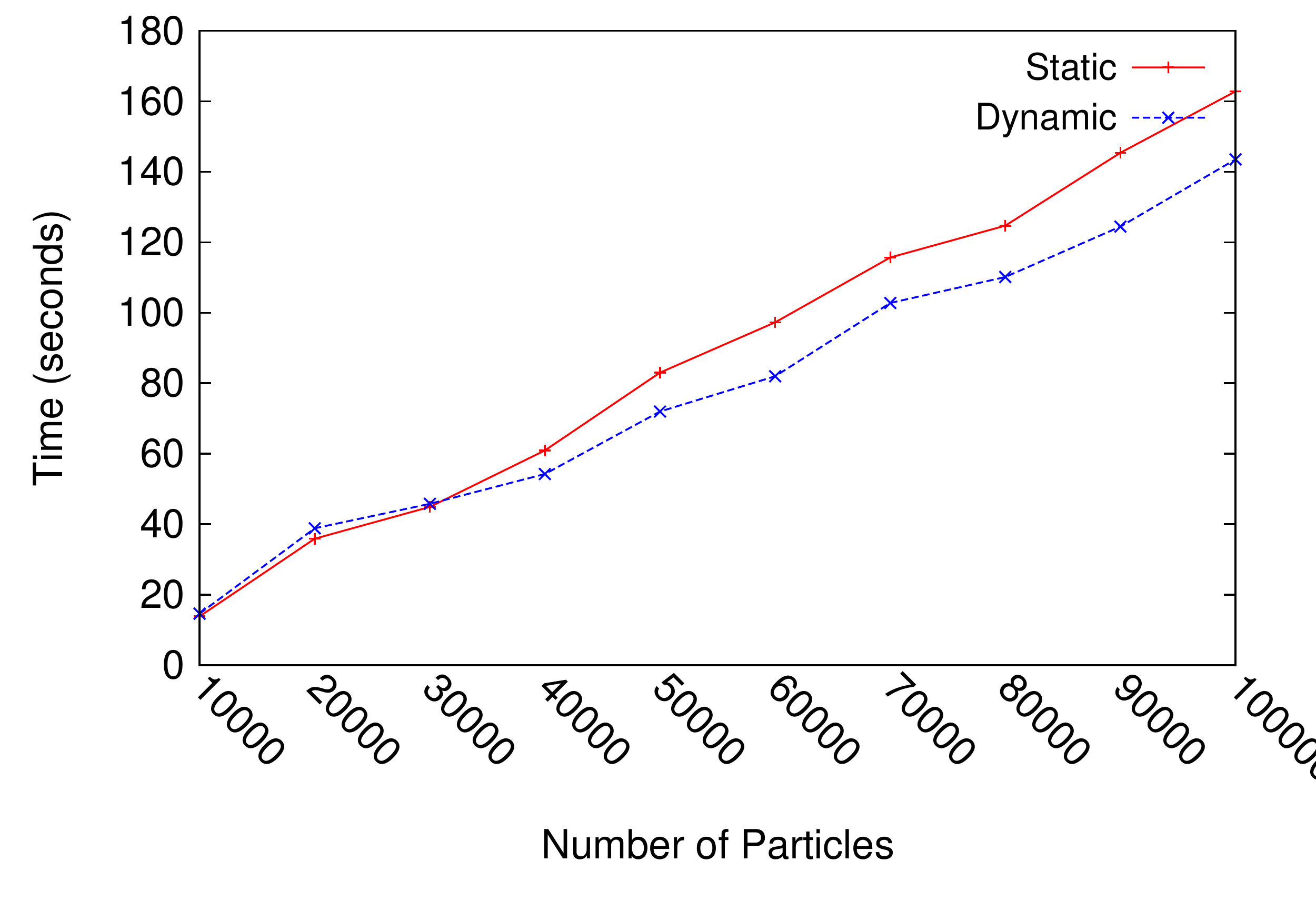}
\caption{Total Execution Times for MD Simulations}
\label{md_total_time}
\end{figure}

Figure \ref{md_total_time} shows the total times taken by the MD simulations with different number of particles using both the static and adaptive strategies for dynamic scheduling.
We find that the adaptive strategy for dynamic scheduling results in 10-15\% reduction in execution times over the static strategy. We also obtained about 22\% reduction in execution time over single-core CPU implementation.

\section{Related Work}
\label{related}

There have been a number of efforts in developing runtime frameworks for efficient executions on GPU systems.

StarPU \cite{augonnet-starpu-cpe2011} is a dynamic task creation and scheduling framework for heterogeneous systems. StarPU's strength lies in its ability to automatically schedule tasks on one or more compute devices. However, StarPU's support for data allocation is completely manual. The programmer has the responsibility of identifying data allocation granularity, task-to-data mapping and inter-task data dependences. These are done automatically in our G-Charm framework.

Our work is closely related to the work by Kunzman \cite{kunzman-runtime-thesisuiuc2012} that has developed a unified programming model for abstracting different types of accelerators, with the runtime system performing various tasks such as load balancing, work agglomeration and data management. In this work, the user has to explicitly specify if a given data should be persistent in GPU memory across kernel invocations to avoid redundant data transfers, while in our work, such data management is performed automatically by the runtime system.  All these efforts do not consider the non-uniform workloads and irregular data access patterns in irregular applications. In our results, we show that adaptive strategies that consider these aspects give significant benefits for irregular applications over strategies that assume regular workloads and access patterns.

\section{Conclusions and Future Work}
\label{con}

In this work, we had developed adaptive strategies for efficient executions of irregular message-driven applications on GPU systems.
By means of experiments with ChANGa N-Body simulations and MD applications, we showed that our dynamic strategies result in 8-38\% reduction in execution times for these irregular applications over the corresponding static strategies that are amenable for regular applications. For the N-Body simulations, we also showed that our generic framework and strategies performs competently with a hand-tuned and optimized code.

\bibliographystyle{abbrv}
\bibliography{strategies-irregular}

\end{document}